\documentclass[referee]{aa} 
\nolinenumbers
\usepackage{graphicx}
\usepackage{txfonts}
\newcommand{\kms}{km\,s$^{-1}$}
\newcommand{\rsun}{R$_\odot$}

\begin{document}

   \title{Formation of a streamer blob via the merger of multiple plasma clumps below 2\,\rsun}
   \titlerunning{Formation of a streamer blob below 2\,\rsun}


 \author{Haiyi Li \inst{1}
          \and
          Zhenghua Huang \inst{1}\fnmsep \thanks{z.huang@sdu.edu.cn}
          \and
          Kaiwen Deng  \inst{1}
          \and
          Hui Fu \inst{1}
          \and
          Lidong Xia   \inst{1}
          \and
          Hongqiang Song  \inst{1}
          \and
          Ming Xiong \inst{2}
          \and
          Hengyuan Wei \inst{1}
          \and
          Youqian Qi \inst{1}
          \and
          Chao Zhang \inst{1}
          }

   \institute{Shandong Key Laboratory of Optical Astronomy and Solar-Terrestrial Environment, Institute of Space Sciences,
             180 Wenhua Xilu, Weihai, 264209, Shandong, China\\
         \and
            Key Laboratory of Solar Activity and Space Weather, National Space Science Center, Chinese Academy of Sciences, Beijing, 100049, China\\
             }

\authorrunning{Li H. et al.}

   \date{Received ???; accepted ???}

 
  \abstract
   {Propagating streamer blobs could be an important source of disturbances in the solar wind.
Direct observations on formation of streamer blobs  could be a proxy for understanding the formation of small-scale structures and disturbances in the solar wind.}
   {We aim to investigate how a streamer blob is formed before it is observed in the outer corona.}
   {Using special coordinated-observations from SOHO/LASCO, GOES/SUVI and SDO/AIA, we study the precursors of a streamer blob as seen in the corona below 2.0 solar radii (\rsun). 
}
   {We found that the streamer blob was formed due to the gradual merging of three clumps of brightenings initiated from the lower corona at about 1.8\,\rsun, which is likely driven by expansion of the loop system at the base of the streamer. 
  The acceleration of the blob starts from 1.9\,\rsun\ or lower. 
 It propagates along the south flank of the streamer where an expanding elongated brightening occurs coincidently.
 }
   {Our observations demonstrate that formation of a streamer blob is a complex process. We suggest that the expansion of the loop  results in a pinching-off flux-rope-like blob at the loop apex below 2\,\rsun.
When the blob moves outward, it can be transferred across the overlying loops through interchange/component magnetic reconnection and then is released into the open field system.
When the blob moves toward open field lines, interchange magnetic reconnections might also occur, and that can accelerate the plasma blob intermittently whilst allow it to transfer across the open field lines.
Such dynamics in a streamer blob might further trigger small-scale disturbances in the solar wind such as switchbacks in the inner heliosphere.
}


   \maketitle
%

\section{Introduction} \label{sec:intro}
Coronal streamers are large-scale quasi-static structures rising above filaments and/or filament channels,which are extending from the solar surface to the interplanetary space for many solar radii and are one of the most magnificent features seen during a solar eclipse\,\citep[e.g.][]{1973SoPh...31..105S,2009SoPh..258..243A,2011ApJ...734..114P,2015ApJ...800...90P,2020ApJ...895..123B,2023MNRAS.518.1776L}.
They are believed to be the results of a complex interaction between the surrounding solar wind and large-scale magnetic fields\,\citep{1992ApJ...392..310W}. 
They can root at almost all latitudes during a sunspot maximum, and concentrate around the equator during a sunspot minimum\,\citep{2007A&A...475..707S,2005A&A...430..701V}.
The formation of a current system with the slow dissipation of the magnetic field is thought to be responsible for maintaining such a stable structure over many solar radii and several months\,\citep{1992SSRv...61..393K}.
There are two classes of coronal streamers, helmet streamers and pseudo-streamers\,\citep{2007ApJ...658.1340W}.
Helmet streamers consist of systems of magnetic loops and the open magnetic field lines with opposite magnetic polarities separated by the loops\,\citep{1968SoPh....5...87S, 2021A&A...656A..32R}.
The plasma escaping along these open magnetic field lines forms the helmet streamer rays\,\citep{1999SoPh..188..299E, 2020ApJS..246...60P}.
The polarity inversion line or coronal neutral line that forms near the top of helmet streamers is the coronal origin of the heliospheric current sheet\,\citep{1993JGR....98.9371C,2009JGRA..114.4103S}.
Helmet streamer rays are thought to engulf the heliospheric current sheet (HCS) and be the source of the heliospheric plasma sheet (HPS) typically measured in-situ while crossing the HCS\,\citep{1994JGR....99.6667W,2000A&A...362..342G,2020ApJS..246...37R}.

\par
Coronal streamers are dynamic, in which small-scale transients such as blobs frequently occur\,\citep{1997ApJ...484..472S,1998ApJ...498L.165W,2009SoPh..258..129S,2009ApJ...694.1471S,2012SoPh..276..261S,2023A&A...672A.100L}. 
Such dynamics in coronal streamers might be related to formation of slow solar wind detected in the interplanetary space, including that near the earth\,\citep{2000A&A...357.1051L,2021A&A...656A..32R,2023A&A...675A.170V}.
By comparing the measured outflow velocities and abundances of similar elements with the slow solar wind measured in-situ, many previous studies have concluded that the legs and stalks of the coronal helmet streamer may be the source regions of the slow solar wind\,\citep{1997SoPh..175..645R,2002ApJ...571.1008S,2003ApJ...585.1062U,2021A&A...650A..12Z,2021A&A...650L...3C}. 

\par
\citet{1997ApJ...484..472S} studied white-light coronal images taken by SOHO/LASCO and they traced small scale non-uniform structures in coronal helmet streamer with a length of about 1\,\rsun (solar radius) and a width of about 0.1\,R$_\odot$.
These structures were found to move radially outward to a distance of 3-4\,\rsun\ at least, and referred to as `blobs' due to their density inhomogeneity.
Using data from the twin satellites of STEREO, \citet{2009ApJ...694.1471S} analysed streamer blobs observed simultaneously with edge-on and face-on views and found that the blobs have a concave-outward structure in the edge-on views and an arch topology in the face-on views.
This suggests a flux-rope nature of streamer blobs as proposed in simulations\,\citep{2006ApJ...642..523T} and is consistent with observations of gradually accelerating CMEs related to streamer blobs\,\citep{2006ApJ...650.1172W,2007ApJ...655.1142S}.
\citet{2004JGRA..109.3107C} concluded the materials in blobs and coronal streamers are confined to narrow sheets of plasma, and blobs may be the counterpart of the heliospheric plasma sheet with high $\beta$ value. 
By tracing them in coronal images taken by SOHO/LASCO, these blobs are found to be accelerated while they are propagating away from the sun, and found to be in a range below 150\,\kms\ at a distance of 3--4\,\rsun and about 400\,\kms\ at a distance of about 25\,\rsun\,\citep{1997ApJ...484..472S,1998ApJ...498L.165W,1999JGR...10424739S}.
The latter one is characteristic of slow solar wind speed, and thus these blobs are a candidate of the sources of slow solar wind\,\citep{1997ApJ...484..472S,1998ApJ...498L.165W,1999JGR...10424739S}.
Such a proposal is also demonstrated by a recent work by \citet{2020ApJS..246...37R}, in which they trace a series of coronal streamer blobs from about 4\,\rsun\ to about 100\,\rsun\ in the observations of STEREO-A/COR2/HI1, and they found those streamer blobs are corresponding to enhanced density structures lasting from tens of minutes to tens of hours at about 40\,\rsun\ in the in-situ observations from PSP.

\par
The origin of streamer blobs is thought to be crucial because that might directly link to the origin of solar wind. 
\citet{1998ApJ...498L.165W} proposed a scenario that interchange reconnection between loops and open field lines could release materials trapped in the close fields into open fields and then form the observed blobs in the coronal streamers (see their Fig. 8).
Simulations of interchange reconnection dynamics in a solar pseudo-streamer has revealed that coronal plasma with closed field can be injected into the interplanetary space\,\citep{2023A&A...675A..55P}.
Alternatively, the simulation carried out by \citet{2009ApJ...691.1936C} suggests that the formation of blobs could be results of an intrinsic instability of coronal streamers around the cusp due to the dynamic equilibrium between the weak magnetic field constraint and the high temperature expansion of the plasma.
In an MHD simulation, \citet{2018ApJ...859....6H} found that the flux rope structures of a blob in a coronal helmet streamer could be generated by pinching-off reconnections as the streamer top stretching out.

\par
To understand what really happen in responsible for formation of a streamer blob, observations of its origins are crucial.
Recently, \citet{2021ApJ...920L...6L} used coordinated observations from ground-based coronagraph K-Cor and LASCO C2 to study the dynamics of streamers in great details.
They found that blobs are generated along the centers of helmet streamers and pseudo-streamers and also their legs and finally merge into their stalks.
They also found that blobs formed along the centers of the helmet streamers origin at average heights of about 2.6\,\rsun\ that is below the top of the streamer cores,
while blobs formed along the legs of helmet streamers and pseudo-streamers might be generated at or below 2\,\rsun.
Unfortunately, in their study, streamer blobs were not seen below 2\,\rsun\ by K-Cor possibly due to the limitation of ground-based observations, and thus the forming processes of streamer blobs at very beginning remain unclear.

\par
Specially-designed observations from the Solar Ultraviolet Imager\,\citep[SUVI,][]{2018ApJ...852L...9S,2019SPIE11180E..7PV} on the Geostationary Operational Environmental Satellite 17 (GOES-17) spacecraft have revealed very dynamic natures of the middle corona in a range of 1.5--3\,\rsun, providing evidence that the solar wind structures, including those carried by streamer blobs, have origins in this region\,\citep{2021NatAs...5.1029S,2023NatAs...7..133C,2023SoPh..298...78W}.
Here, we make use of these specially-designed data, in combination of the data from the C2 coronagraph of the Large Angle and Spectrometric Coronagraph\,\citep[LASCO C2,][]{1995SoPh..162..357B} on the Solar and Heliospheric Observatory\,\citep[SOHO,][]{1995SoPh..162....1D} and the Atmospheric Imaging Assembly\,\citep[AIA,][]{2012SoPh..275...17L} on the Solar Dynamics Observatory\,\citep[SDO,][]{2012SoPh..275....3P}, we investigate in-depth the formation processes of a streamer blob.
We aim to find out from the very beginning where the streamer blob is formed and what activities in the lower corona are responsible,
and from that we make further discussion on the mechanism of its formation.
In what follows, we give descriptions of the analysed data in Section\,\ref{sec:obs}, the results in Section\,\ref{sec:res}, the discussion in Section\,\ref{sec:dis} and conclusions in Section\,\ref{sec:con}.

\begin{figure*}[!ht]
\centering
\includegraphics[width=\textwidth]{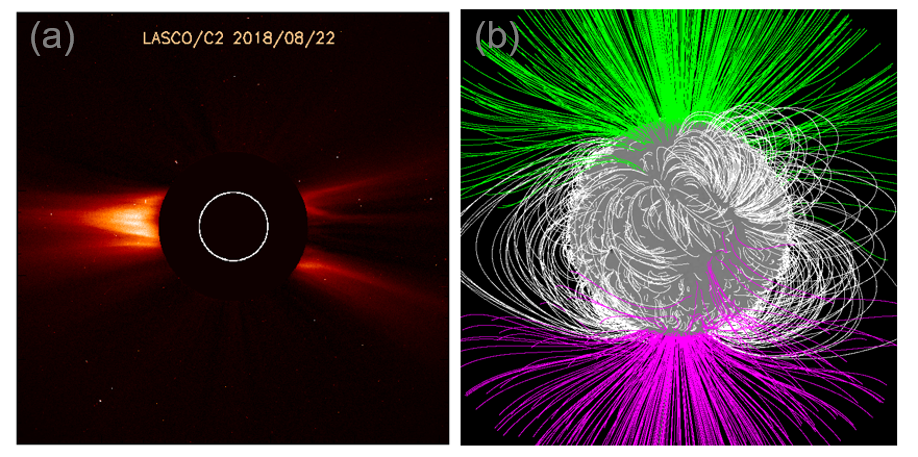}
\centering
\caption{(a): polarized white-light image of the corona recorded with the LASCO C2 on 2018 August 22.
The inner boundary of the field-of-view is at 2\,\rsun, and the outer boundary is at about 6\,\rsun.
The white circle indicates the limb of the solar disk.
(b): potential magnetic field extrapolated with PFSS model overlying on a photospheric synoptic magnetogram.
The purple and green lines are representative of open field lines and white lines are closed loops.
\label{fig:1}}
\end{figure*}

\begin{figure*}[!ht]
\centering
\includegraphics[width=\textwidth]{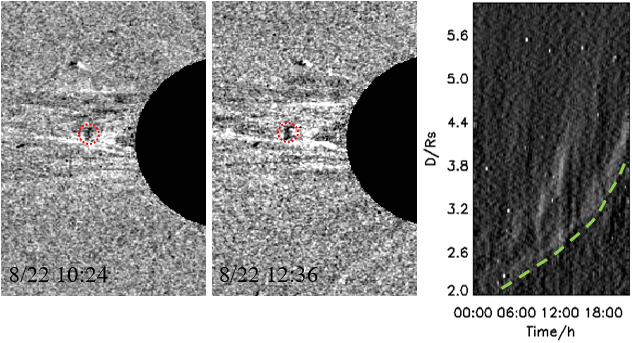}
\caption {Left and middle panels: running-difference images obtained from LASCO C2 observations at 10:24\,UT and 12:36\,UT on August 22, on which a small propagating blob is shown as a structure with dark center surrounded by bright features (marked by circles in red dashed lines). 
Right panel: a time-distance map tracking the propagation of the streamer blob on the running difference images.
The time starts from 00:00\,UT on August 22.
The dashed line in green denotes the propagation of the streamer blob, from which the tangent derives its propagating speeds.
An associated animation is given online.
\label{fig:2}}
\end{figure*}

\begin{figure*}[!ht]
\centering
\includegraphics[width=\textwidth]{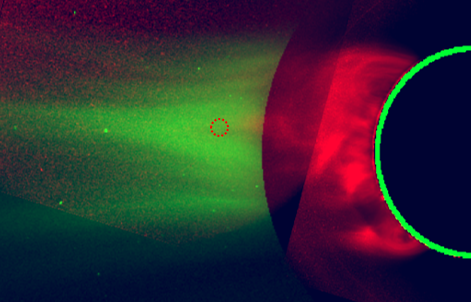}
\caption {The streamer observed in the compositional images of SUVI 193\,\AA\ (red) and LASCO C2 (green).
The red small circles indicate the location of the blobs recognized in LASCO C2 running-difference images (i.e. left and middle panels of Fig.\,\ref{fig:2}). An associated animation is given online.
\label{fig:3}}
\end{figure*}

\begin{figure*}[!ht]
\centering
\includegraphics[scale=0.99]{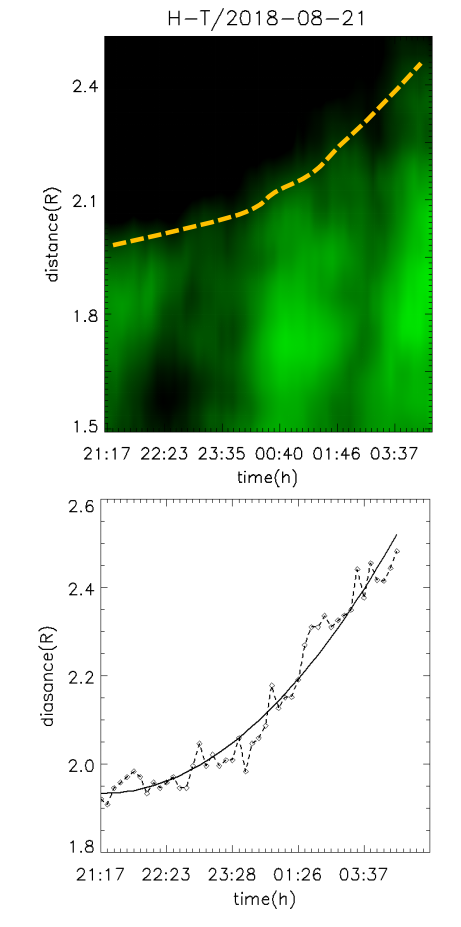}
\caption{Top: time-distance map following the expansion of the elongated bright feature in the SUVI images (the time coordinate starts from August 21 21:17\,UT).
The blob presents as the tip of the expanding feature, which can be tracked by the gold dashed line. 
Bottom: the variation of the heights of the blob seen in SUVI images (diamonds with dashed line), which are identified manually.
The solid line is a second-order polynomial fit to the variation curve.
\label{fig:4}}
\end{figure*}

\begin{figure*}[!ht]
\centering
\includegraphics[width=\textwidth]{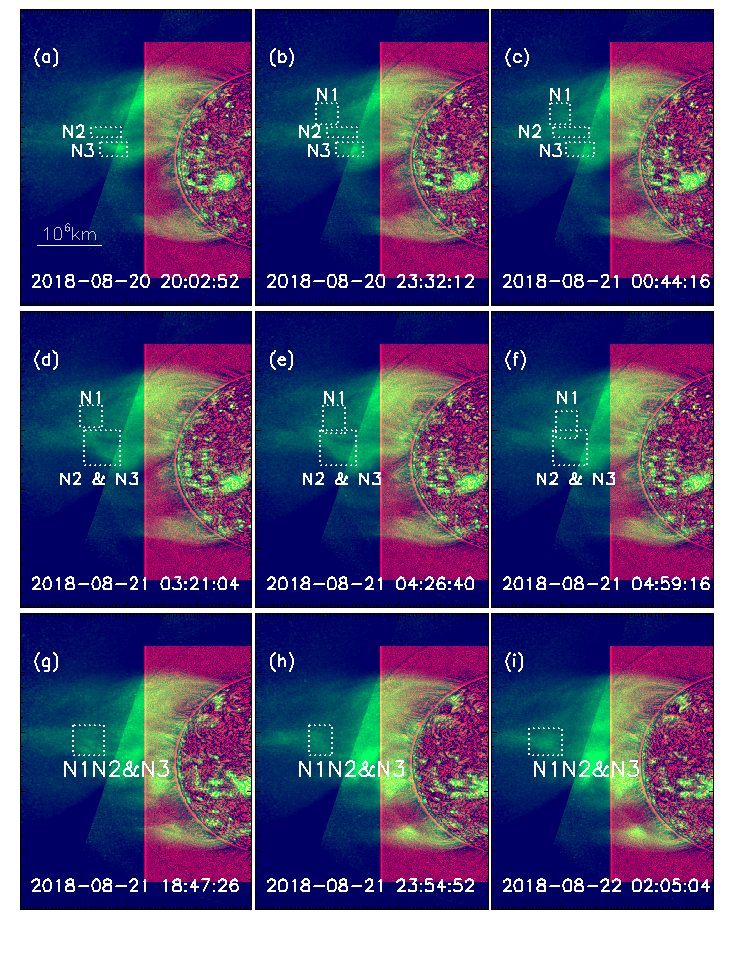}
\caption{Evolution of the solar corona seen in combined observations of AIA 193\,\AA\ (red) and SUVI 195\,\AA\ (green).
The regions enclosed by dotted lines denote three clumps of brightenings (N1, N2 and N3), and their merging processes can be followed.
\label{fig:5}}
\end{figure*}

\begin{figure*}[ht!]
\centering
\includegraphics[width=\textwidth]{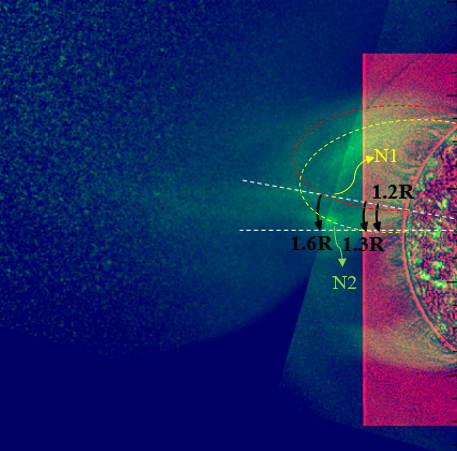}
\caption{A combined image of SUVI 195\,\AA\ image and AIA\,193\,\AA\ image.
The red and yellow dashed lines mark the loops associated with N1 and N2, respectively.
Three slits at 1.2\,\rsun, 1.3\,\rsun and 1.6\,\rsun where we obtain distance-time maps (shown in Figure\,\ref{fig:8}) are marked as black arrows (0 is given at the tail end).
These slits are arranged roughly in the same opening angle.
Please note that the elongated structures are not perfectly aligned in the radial direction.
An associated animation is given online.
\label{fig:6}}
\end{figure*}

\begin{figure*}[ht!]
\centering
\includegraphics[width=\textwidth]{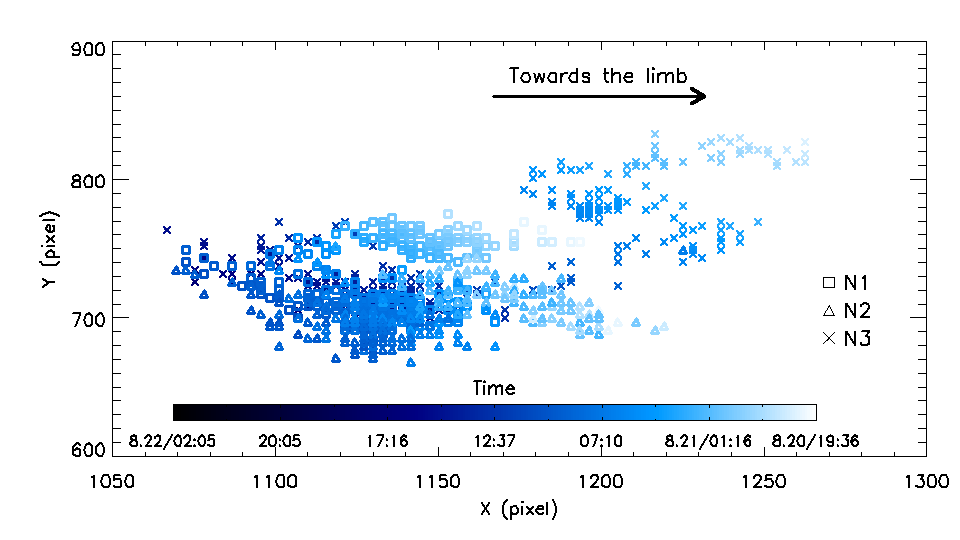}
\caption{Variations of centers (on the projected X-Y plane) of the three clumps of brightenings.
The squares, triangles and crosses are locations of ``N1'', ``N2'' and ``N3'', respectively.
The changing in time is shown in variation of color.
The convergence of the three clumps can be followed in this diagram while they are moving toward the same position at the end. 
\label{fig:7}}
\end{figure*}

\begin{figure*}[ht!]
\centering
\includegraphics[trim=1cm 0cm 1cm 0cm,width=\textwidth]
{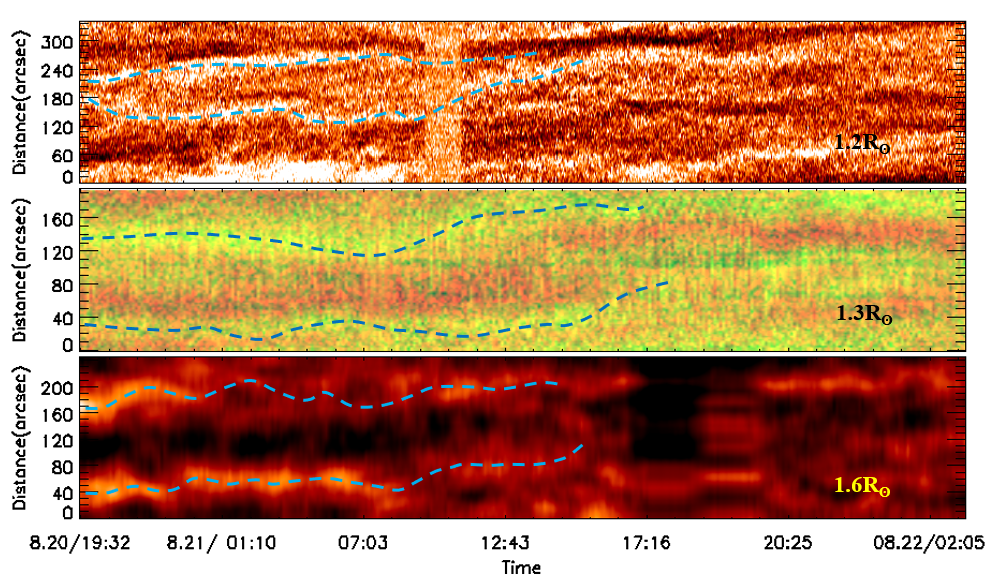}
\caption{Time-distance maps obtained from the three slits shown in Figure\,\ref{fig:6} based on AIA observations (1.2\,\rsun) and SUVI observations (1.3\,\rsun\ and 1.6\,\rsun).
The swaying motion of the extending structures (in which ``N1'' and ``N2'' are sitting in) at 1.2, 1.3 and 1.6\,\rsun\ can be tracked with the dashed lines.
\label{fig:8}}
\end{figure*}
\section{Observations} \label{sec:obs}
The coordinated data analysed in the present study were taken by GOES-17/SUVI, 
SOHO/LASCO C2 and SDO/AIA.
GOES-17/SUVI is an EUV imager equipped with a CCD with 1280$\times$1280 pixels and a pixel scale of 2.5 arcsecond per pixel.
LASCO C2 is an externally-occulted white-light coronagraph with a field-of-view extending from 2\,\rsun\ to 6\,\rsun, a pixel size of 11.4\,arcsecond and a cadence of 45\,s.
SDO/AIA is a set of EUV imagers observing the full disk of the sun with a resolution of 0.6 arcsecond per pixel and a cadence of 12\,s.

\par
We analysed the LASCO C2 data taken in the period between 00:00 UT to 23:59 UT on 2018 August 22, during which we observer a streamer blob initiating from the location of 2 \rsun.
The data have been calibrated by standard procedures provided by the instrument team along with {\it solarsoft}.

\par
The SUVI data were specially-designed to obtain the EUV images of the Sun with an east–west rastering mode so that the FOV can extend to about 5 \,\rsun\ in the horizontal direction.
In this observing mode, it cannot provide simultaneous views of the disk while pointing to the side.
For each pointing, the cadence of the observations is around 20 minutes but not regular.
More details of these observations can be found in \citet{2021NatAs...5.1029S}.
The SUVI data have been calibrated by the instrument team and no further calibration is required.
In order to trace the evolution of the inner corona below 1.3 \,\rsun, we use observations from AIA 193\,\AA\ passband,
and in coordination with the SUVI images taken at the passband of 195\,\AA.
These two passbands observe the solar structures with similar representative temperatures, and the streamer structures are best seen.
The AIA data are calibrated by the standard procedures provided by the instrument team.

\par
In order to see better the coronal structures in the images, several techniques are employed.
For the LASCO C2 images, the time-independent F corona is removed by subtracting out a time-averaged background and then running different images are produced for bringing out the faint and moving blob in the streamer.
For SUVI images, we first apply a Normalizing Radial Graded Filter\,\citep[NRGF,][]{2006SoPh..236..263M} and then a two-dimensional Savitzky-Golay filter with 7$\times$7 window, from that we can reveal more details of the streamer\,\citep{2021NatAs...5.1029S}.
The data from different instruments are co-aligned by cross-checking the connections of the global structures near the limb, 
and we can see a good correlation of the coronal structures from low to middle corona. 
To trace the evolution of the activities in the streamer appearing at the east limb, the center part of a SUVI image is replaced by the AIA image at the near time as that of the east-side-pointing image.



\section{Results}\label{sec:res} 
The image of the corona seen by LASCO C2 is shown in panel (a) of Figure\,\ref{fig:1}, while the potential field given by the PFSS model is shown in panel (b).
We can see that the streamer is highly flattened toward the heliographic equator because the year of 2018 is near the solar minimum\,\citep{2000GeoRL..27..149W}, which is consistent with the PFSS model.
We concentrate on the the streamer appearring at the east limb,
which can be considered as helmet streamer since it separates open field lines with opposite polarities as shown in the PFSS model.
It consists of a cluster of coronal loops and open field lines at both sides.

\par
To follow the evolution of the streamer, the running different images from LASCO C2 observations are produced (see Figure\,\ref{fig:2}).
A small blob in the streamer in the LASCO C2 field-of-view is firstly seen at 02:20\,UT on 2018 August 22 (see the structure denoted by red circles in the left and middle panels of Figure\,\ref{fig:2}).
The blob has a size of about 0.4\,\rsun, which is typical in these phenomena\,\citep{1997ApJ...484..472S,2009ApJ...694.1471S}.
The blob propagates slightly toward the south flank of the streamer, where we also observe expansion of a narrow elongated bright feature.
The expansion of the narrow bright feature is conincident with the propagation of the blob.
Their possible connection will be discussed further later.
We then make a height-time (H-T) plot to follow the propagation of the blob as shown in the right panel of Figure\,\ref{fig:2}.
From the H-T plot, we can obtain the propagation speeds of the blob from the slopes of the tangents of the trajectory.
We can see that  the blob is accelerated from about 14\,\kms\ at 2\,\rsun\ to about 42\,\kms\ at 4.5\,\rsun, which agrees with previous studies\,\citep{2000GeoRL..27..149W}.

\par

The connection of the blob observed by SUVI and LASCO/C2 is shown in Fig.\,\ref{fig:3} and the associated animation.
We can see the overall structures of the streamer seen in SUVI and whitelight are well consistent.
In the animation, we can see that the blob formed at about 1.8\,\rsun\ and gradually moved radially outward. 
The blob apparently formed at the tip of streamer loops as seen in the SUVI images. 
While the streamer loops expand, we observe the leading edge of the blob first becomes sharp and thin, and then moves outward. 
It reaches 2.0\,\rsun\ at about 00:00\,UT on August 22.

\par
In Fig.\,\ref{fig:4}, we show time-distance diagrams of the blob based on the SUVI observations.
We found that the speeds of the blob are ranging from $\sim8$\,\kms\ (at 1.9\,\rsun) to $\sim25$\,\kms (at 2.5\,\rsun).
Such dynamics of the blob viewed in SUVI is consistent with that seen in LASCO/C2.
We can see that the curve of the distances (Fig.\,\ref{fig:4}) can be well fitted by a second order polynomial function.
This suggests that acceleration of the blob starts from 1.9\,\rsun\ or lower.
Together with the whitelight observations (Fig.\,\ref{fig:2}), we can see that the streamer blob is continuously accelerated for several solar radii.
The mechanism driving such an acceleration should be an important proxy for understanding the formation of solar wind.

We then trace the origin of the blob back in time and in lower corona as seen in the SUVI and AIA observations.
In Figure\,\ref{fig:5}, we show the evolution of the solar corona in the combined SUVI and AIA observations.
We found that the blob is formed via convergence of three clumps of brightenings.
The first two clumps are first seen as early as 20:02\,UT on August 20 (see the features in the boxes denoted by ``N1'' and ``N2'' in Figure\,\ref{fig:5}\,(a)).
Both N1 and N2 are elongated structures extending more-or-less radially, apparently located in the south flank of the streamer.
Due to their bright natures against the ambient corona, they are likely parts of the south legs of the outer layer of the loop system (see the red and yellow dashed lines in Figure\,\ref{fig:6}), which is also consistent with the field extrapolation shown in Figure\,\ref{fig:1}.

N1 has about 0.1\,\rsun\ in length and 0.02\,\rsun\ in width, while N2 has about 0.12\,\rsun\ in length and 0.05\,\rsun\ in width.
These values are not very accurate because they are faint and are not much outstanding. 
N1 starts at about 1.7\,\rsun\ and N2 starts slightly lower at about 1.6\,\rsun.
At about 23:32\,UT on August 20, another clump (N3) appears to be pinching-off at the top of the cluster of coronal loops ($\sim1.7$ \rsun) in the streamer base and to the north of N1 and N2 (see Figure\,\ref{fig:5}\,(b)\&(c)).
These loops seem to be low-lying ones underneath those hosting N1 and N2.
The newly formed N3 is apparently approaching N1, and we observed that N1 and N2 are moving outward and begin to merge at the time around 03:21\,UT on August 21 (see Figure\,\ref{fig:5}\,(d)).
The convergence of N1 and N2 occurs at a height of about 1.8\,\rsun\ (see Figure\,\ref{fig:5}\,(d)--(f)).
Meanwhile, N3 keeps moving toward the top point of the convergence of N1 and N2, and they begin to merge around 18:47\,UT on August 21 (Figure\,\ref{fig:5}\,(g)).
During the next eight hours, the merged structure becomes more compact at the height of about 2\,\rsun, and then starts moving outward and forms the blob seen in the LASCO C2 FOV (Figure\,\ref{fig:5}\,(g)--(i)).

\par
We also track these convergence processes manually to identify the variations of the centers of the three clumps.
In Figure\,\ref{fig:7}, we show variations of their coordinates on the heliographic projection plane.
Along X direction N1 and N2 are moving away from the limb unidirectionally,
while alongY direction they show some fluctuations.
N3 shows forth and back motions along X, but along Y it moves southward at most of the times.
The effective speeds of N1, N2 and N3 before their convergence are 1.5\,\kms, 0.6\,\kms\ and 0.7\,\kms\, respectively.
 However, these merging processes are not taken into account the projection effect, and the real speeds could be much larger.

\begin{figure*}[ht!]
\centering
\includegraphics[width=\textwidth]{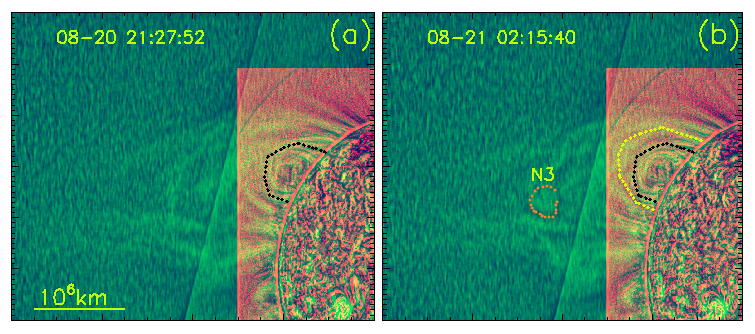}
\caption{The birth of ``N3'' (enclosed by red dotted line in panel\,(b)) and expansion of low-lying loops seen in the combined SUVI and AIA images. 
The black dotted lines denote the location of the loop seen in the AIA image at 21:27\,UT on August 20 and the yellow dotted line marks that at 02:15\,UT on August 21. 
An associated animation is given online.
\label{fig:9}}
\end{figure*}

\begin{figure*}[ht!]
\centering
\includegraphics[width=0.8\linewidth]{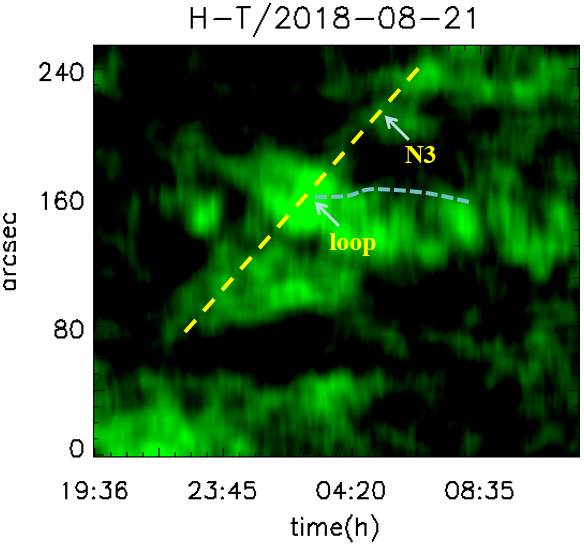}
\caption{Time--distance map obtained along the expanding direction of the loop system and the propagation of ``N3''.
The blue arrow indicate the time and location when and where ``N3'' is pinching-off.
The yellow dashed line shows the expansion of the loop system and the falling-off of ``N3''.
The cyan dashed line indicates the retraction of the loop system.}
\label{fig:10}
\end{figure*}

\par
Then, we check carefully the origins of these clumps of brightenings.
Although N1 and N2 appear gradually without any clear dynamic processes,
we see that the threads where N1 and N2 extend along show swaying motions in the south-north direction before the two clumps moving outward.
Such a motion can be better seen in the online animation associated with Figure\,\ref{fig:6}, in which the locations of the swaying threads are indicated by the red arrows and the final combination of the three clumps are denoted by the yellow arrows.
Such swaying motions can also be indicated in the fluctuations of their Y coordinates shown in Figure\,\ref{fig:7}.
We further take three slices perpendicular to the threads at 1.2\,\rsun, 1.3\,\rsun\ and 1.6\,\rsun\ (Figure\,\ref{fig:6}) to produce space-time maps (see Figure\,\ref{fig:8}).
The swaying motions of the two threads at these heights seems to be not in sync.
We find that N3 is born at the apex of loop-like structures, likely with a pinching-off process (see the region marked in Figure\,\ref{fig:9}).
At the beginning (see e.g. Figure\,\ref{fig:9}\,(a)), we see the loop-like structures in the region is sparse.
In contrast, it becomes more chaotic when N3 is born (see Figure\,\ref{fig:9}).
During this period of time, we also observe that the coronal loops in the lower corona (AIA FOV) show a clear expansion from about 1.4\,\rsun (the black fotted line in Figure\,\ref{fig:9}\,(a)\&(b)) to about 1.5\,\rsun\ just prior to the birth of N3 (the golden dotted line in Figure\,\ref{fig:9}\,(b)), which the solar rotation cannot compensate.
Such evolution of the birth of ``N3'' can be followed in the online animation associated with Figure\,\ref{fig:9}.
A time--distance map demonstrating the birth of N3 is shown in Figure\,\ref{fig:10}, from which we can see that the expansion of the loops system has an apparent speed of about 4\,\kms\ (see the yellow dashed line in the figure).
The detachment of  N3 is also accompanied by slightly contraction of the loop system with an apparent speed less than 1\,\kms\  (see the cyan dashed line after in Figure\,\ref{fig:10}).
Such dynamics is similar to those reported previously by \citet{2018ApJ...859..135W}, but the present case occurs at much lower height (below 2\,\rsun) and the speeds are also one to two magnitude smaller.
Although we cannot find any direct connection between this motion of loop expansion and the birth of N3 due to the low resolution of the data, we suspect that the expanding loops might provide driver to trigger the dynamics at the overlying loops and result in instability and/or reconnection there, similar to that suggested by \citet{2018ApJ...859..135W}.
The appearance of N3 and its motion toward the two threads of N1 and N2 seem to be the direct trigger to the convergence of the three clumps and then the formation of the streamer blob.

\begin{figure*}[ht!]
\centering
\includegraphics[width=\textwidth]{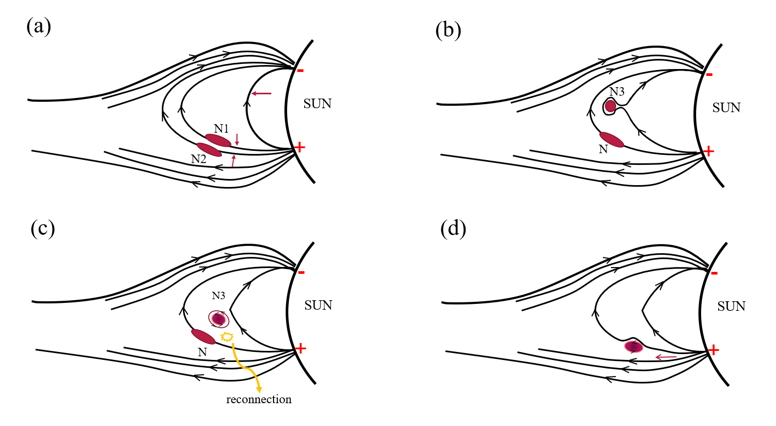}
\caption{A schematic cartoon model of the formation of the streamer blob observed here.
(a): the general magnetic topology of the streamer.
The bright features of ``N1'' and ``N2'' are located in the south legs of the overlying loops.
The horizontal red arrows indicate the upward expansion of the loop system and that results in the squeezing of the overlying loop system (the vertical red arrows).
(b): the geometry of the streamer while ``N1'' and ``N2'' have merged  (or simply been arranged along the same line-of-sight) and pinching-off processes take place near the apex of the lower-lying loops.
(c): the geometry of the streamer when a flux-rope-like feature (``N3'') is fallen-off due to the pinching-off processes.
Magnetic reconnection might take place while the flux-rope moving toward the overlying loop system, allowing it to escape from the close field. 
(d): the geometry of the streamer after that the flux rope is released in the open field system.
The distortion of the field line resulted from the magnetic reconnection could provide additional magnetic tension to drive the blob to move outward.
The flux-rope might also continuously interact with the open field and interchange magnetic reconnection might occur intermittently all the way out until it is fully dissolved.
\label{fig:11}}
\end{figure*}

\section{Discussion}\label{sec:dis}
Based on these observations, we confirm that the formation of a streamer blob can be a complex process. Here we suggest the streamer blob observed here  possibly form via processes as shown in Figure\,\ref{fig:11}.
A possible magnetic system of the streamer is shown in panel\,(a) of Figure\,\ref{fig:11}.
It consists of a loop system, together with open fields rooted at both of its sides.
The clumps of ``N1'' and ``N2'' are coupled with the south legs of the outer loops.
As the low-lying loops expand, they might force the outer loops that host ``N1'' and ``N2'' to converge and the two clumps might merge or simply be arranged at the same location on the projected plane (panel (b)).
Meanwhile, the expanding motion from the below of the loop system might lead to density inhomogeneities and/or  intrinsic instability due to misalignment of the field lines and thus uneven forces at different locations near the loop top, which includes such as tearing mode instability\,\citep{1998ApJ...495..491K,1999JGR...104..521E}, pinching-off reconnection\,\citep{2018ApJ...859....6H} or simply the magnetic diffusion\,\citep{2009ApJ...691.1936C}, and this results in ``N3'' to be broken-down (Figure\,\ref{fig:11}\,(b)\&(c)).
However, any of these processes cannot be demonstrated with the present data due to the limitation of spatial and temporal resolutions.
Since streamers normally consist of filaments in their bases and streamer blobs are mostly consistent with flux-rope geometries\,\citep{2009ApJ...694.1471S}, ``N3'' likely consists of a flux-rope-type magnetic field.
While ``N3'' moves toward the south, it could interact with the outer loops hosting ``N1'' and ``N2'' and might result in magnetic reconnection between the loop systems and the flux-ropes (Figure\,\ref{fig:11}\,(c)).
This can be a kind of component reconnection that allows the flux-rope to escape from the close field and to be released into the open field system (Figure\,\ref{fig:11}\,(d)).
This motion of a flux-rope could also push the overlying loops to expand, as observed in the SUVI images (see the animation associated with Fig.\,\ref{fig:3}).
Such a process might also provide extra momentum of the flux rope toward the flank of the open field system and that might drive interchange reconnection between the tangential flux of the flux-rope and the open field.
This interchange magnetic reconnection can also explain the expansion of the brightening at the south flank of the streamer.
After such an interchange reconnection, the open field should undergo a restoring process that can force the blob to move outward.
While the photospheric motions can drag the footpoints of loops at the bottom of a streamer, leading to expansion of the loop system, it results in shearing of the streamer\,\citep{1995ApJ...438L..45L} and finally the instabilities and/or magnetic reconnections develop and a streamer blob is formed and ejected into the interplanetary space.
We would like to mention that many details of the scenario proposed here requires observations with higher temporal and spatial resolutions to be proved, which cannot be addressed in the data available presently.

\par
Streamer blobs have been suggested to be an important source of the slow solar wind and/or micro-structures in solar wind\,\citep{1997ApJ...484..472S,1998ApJ...498L.165W,1999JGR...10424739S,2012SSRv..172..123W}.
An important phenomenon in solar wind, switchbacks, has attracted many attentions in the community, since the Parker Solar Probe discovers many details in the recent years\,\citep{2019Natur.576..237B}.
Simulations show that low-amplitude Alfv\'en waves can naturally evolve into a turbulent state with features similar to switchbacks\,\citep{2020ApJ...891L...2S}.
Some studies also suggest that switchbacks might link to activities originated in the low corona or even lower\,\citep[e.g.][]{2020ApJS..246...32T, 2021ApJ...920L..31N,2021ApJ...919...60M,2021ApJ...919...96F,2022ApJ...926..138U,2022EGUGA..24.9673H,2023ApJ...945...28R, 2023ApJ...950...65B,2023ApJ...950..157H}.
A plausible mechanism releasing switchbacks is interchange reconnection between close loops and open field lines in the low corona\,\citep{2020ApJ...903....1Z,2021ApJ...917..110L,2023PhPl...30b2905A}.
Numerical experiments have shown that modified interchange reconnection between open funnels and closed loops can produce both jet flows and  Alfv\'enic wave pulses that can account for the observations of switchbacks in the solar wind\,\citep{2021ApJ...913L..14H}.
Further study suggests that switchbacks are a combination effect of intermittent interchange reconnection and torsional Alfv\'enic waves in the inner heliosphere triggered by the repeated ejections of plasmoid\,\citep{2022ApJ...941L..29W}.
Alternatively, flux ropes formed by multi-x-line reconnection between open and closed flux in the corona could also convect outward and be observed as switchbacks in the inner heliosphere\,\citep{2021A&A...650A...2D}.
Recent observations by Solar Orbiter reveal a single large propagating S-shaped vortex at 2-3\,\rsun\ that appears to be manifestation of interchange reconnection between active region loops and bounded open field system and such structure could be the counterpart of a switchback in the low corona\,\citep{2022ApJ...936L..25T}.
Observations and simulations also suggest that switchbacks and slow solar wind streams can be treated as two manifestations of the same interchange reconnection\,\citep{2022ApJ...936L..25T}.
Whether streamer blobs are associated with formation of switchbacks is an interesting question.
Our observations here indicate a complex formation process of a streamer blob below 2\,\rsun, which possibly includes ejection of a flux-rope-like feature along open field lines.
While such a flux-rope propagates along an open field system, it might intermittently interact with open field lines and result in interchange magnetic reconnection between the tangential field and the open field.
These interchange magnetic reconnections could transfer tangential flux into the open field and provide acceleration to the blob.
Such an interchange magnetic reconnection process might take place all the way the streamer blob propagating outward until it is fully dissolved, which is consistent with observations of prevalence of magnetic reconnection in the near-sun heliosphere\,\citep{2021A&A...650A..13P}.
The transferring of tangential flux in the open field system might drive Alfv\'enic wave pulses locally in the near-sun space.
Therefore, the propagation of such a streamer blob may be associated with a switchback or small flux ropes\,\citep{2021A&A...650A..12Z} in the inner heliosphere.

\section{Conclusions} \label{sec:con}
In this paper, we study the coordinated observations taken by SOHO/LASCO, GOES-17/SUVI and SDO/AIA, aiming to track the formation of a propagating streamer blob down to the low corona near the limb.
The streamer consists of closed loop systems and open field systems at its south and north flanks.
We found that the streamer blob in LASCO/C2 field-of-view was formed via convergence of three clumps (``N1'', ``N2'' and ``N3'') of bright features occurring at about 1.8\,\rsun\ as seen in SUVI 195\,\AA\ passband.
In the SUVI 195\,\AA\ observations, ``N1'' and ``N2'' showing as elongated features at 1.6--1.7\,\rsun\ near the south flank of the streamer appear about 30 hours before the formation of the streamer blob in LASCO C2 images, and ``N3'' forms at the apex of the streamer loop system ($\sim$1.7\,\rsun) about 27 hours before the formation of the streamer blob in LASCO C2 images. 
The convergence of ``N1'' and ``N2'' takes place about 23 hours before the appearance of the streamer blob and it is apparently driven by swaying motions of their field system.
While ``N3'' is fallen from the apex of the loop system, it moves toward ``N1'' and ``N2'' with an effective speed of about 0.7\,\kms\ and merges with them $\sim$8 hours before the appearance of the streamer blob above 2\,\rsun.
The merged clump becomes brighter and more compact, moves outward from about 1.8\,\rsun\ to 2\,\rsun\ in about 8 hours and forms the propagating blob observed in the LASCO C2 field-of-view.
The acceleration of the blob starts at 1.9\,\rsun\ or even lower.
The lower-lying loops in the base of the streamer seen in the AIA field-of-view apparently expand from 1.4\,\rsun\ to 1.5\,\rsun, leading to an expanding speed of about 4\,\kms\ seen in the SUVI images.
When the low-lying loops expand, the field lines hosting ``N1'' and ``N2'' sway and ``N3'' is born at the loop apex.
We speculate that the expansion of the low-lying loops might force the loop systems nearby (``N1'' and ``N2'') to sway, interact and merge, and the expanding loop system itself can lead to inhomogeneity at its apex that further develop into pinching-off ``N3''.
The observations here provide unique evidence on the connection between formation of a streamer blob and activities in the corona below 2\,\rsun.

\par
Based on the observations, we demonstrate that formation of a streamer blob is a complex process, and a scenario for the formation of the observed streamer blob is proposed.
In our scenario, a magnetic flux-rope might be pinching-off from the apex of the low-lying loops due to expansion of the lower loops, which play a crucial role in the formation of the streamer blob.
The fallen-off magnetic flux-rope might interact and/or merge with nearby field systems and that allows it to be transferred across the close field systems and to be released into the open field.
While a flux-rope propagates along open field lines, it might result in intermittent interchange magnetic reconnections, which allow the tangential flux to be transferred across the field lines and provide additional acceleration to the plasma blob.
Our obsrvations suggest that small-scale disturbances in the solar wind observed in the inner heliosphere can be sourced in the corona below 2\,\rsun, but the acceleration processes might take place intermittently on the way it propagates outward.
Our observations demonstrate that a streamer blob itself might consist of multiple sub-structures and thus provide more complexities in the interplanetary space plasma.
The data from the Full Sun Imager (FSI) channel of the Extreme Ultraviolet Imager \,\citep[EUI,][]{2020A&A...642A...8R} aboard the Solar Orbiter\,\citep[SO, ][]{2020A&A...642A...1M} that might cover the corona up to 7\,\rsun\,\citep{2023A&A...674A.127A} should be very valuable for further studies to reveal more details of the formation and dynamics of streamer blobs and their connections to the solar wind.

\begin{acknowledgements}
We are grateful to the anonymous referee for the constructive comments and suggestions.
We thank Prof. Jiansen He for his careful reading to and helpful comments on the manuscript .
We acknowledge the GOES-17/SUVI team for making the data publicly available.
The AIA and HMI data are used by courtesy of NASA/SDO, the AIA and HMI teams and JSOC.
SOHO is a project of international cooperation between ESA and NASA.
The SOHO/LASCO data used here are produced by a consortium of the Naval Research Laboratory (USA), Max-Planck Institute for Sonnensystemforschung (Germany), Laboratoire d'Astrophysique de Marseille (France), and the University of Birmingham (UK).
This research is supported by National Key R\&D Program of China No. 2021YFA0718600 and National Natural Science Foundation of China (42174201, 42230203, 41974201, 42074208). 
\end{acknowledgements}

\bibliography{streamerbib}{}

\begin{thebibliography}{80}
\expandafter\ifx\csname natexlab\endcsname\relax\def\natexlab#1{#1}\fi

\bibitem[{{Ambro{\v{z}}} {et~al.}(2009){Ambro{\v{z}}}, {Druckm{\"u}ller},
  {Galal}, \& {Hamid}}]{2009SoPh..258..243A}
{Ambro{\v{z}}}, P., {Druckm{\"u}ller}, M., {Galal}, A.~A., \& {Hamid}, R.~H.
  2009, \solphys, 258, 243

\bibitem[{{Antonucci} {et~al.}(2023){Antonucci}, {Downs}, {Capuano}, {Spadaro},
  {Susino}, {Telloni}, {Andretta}, {Da Deppo}, {De Leo}, {Fineschi},
  {Frassetto}, {Landini}, {Naletto}, {Nicolini}, {Pancrazzi}, {Romoli},
  {Stangalini}, {Teriaca}, \& {Uslenghi}}]{2023PhPl...30b2905A}
{Antonucci}, E., {Downs}, C., {Capuano}, G.~E., {et~al.} 2023, Physics of
  Plasmas, 30, 022905

\bibitem[{{Auch{\`e}re} {et~al.}(2023){Auch{\`e}re}, {Berghmans}, {Dumesnil},
  {Halain}, {Mercier}, {Rochus}, {Delmotte}, {Fran{\c{c}}ois}, {Hermans},
  {Hervier}, {Kraaikamp}, {Meltchakov}, {Morinaud}, {Philippon}, {Smith},
  {Stegen}, {Verbeeck}, {Zhang}, {Andretta}, {Abbo}, {Buchlin}, {Frassati},
  {Gissot}, {Gyo}, {Harra}, {Jerse}, {Landini}, {Mierla}, {Nicula}, {Parenti},
  {Renotte}, {Romoli}, {Russano}, {Sasso}, {Sch{\"u}hle}, {Schmutz},
  {Soubri{\'e}}, {Susino}, {Teriaca}, {West}, \&
  {Zhukov}}]{2023A&A...674A.127A}
{Auch{\`e}re}, F., {Berghmans}, D., {Dumesnil}, C., {et~al.} 2023, \aap, 674,
  A127

\bibitem[{{Baker} {et~al.}(2023){Baker}, {D{\'e}moulin}, {Yardley},
  {Mihailescu}, {van Driel-Gesztelyi}, {D'Amicis}, {Long}, {To}, {Owen},
  {Horbury}, {Brooks}, {Perrone}, {French}, {James}, {Janvier}, {Matthews},
  {Stangalini}, {Valori}, {Smith}, {Cuadrado}, {Peter}, {Schuehle}, {Harra},
  {Barczynski}, {Berghmans}, {Zhukov}, {Rodriguez}, \&
  {Verbeeck}}]{2023ApJ...950...65B}
{Baker}, D., {D{\'e}moulin}, P., {Yardley}, S.~L., {et~al.} 2023, \apj, 950, 65

\bibitem[{{Bale} {et~al.}(2019){Bale}, {Badman}, {Bonnell}, {Bowen}, {Burgess},
  {Case}, {Cattell}, {Chandran}, {Chaston}, {Chen}, {Drake}, {de Wit},
  {Eastwood}, {Ergun}, {Farrell}, {Fong}, {Goetz}, {Goldstein}, {Goodrich},
  {Harvey}, {Horbury}, {Howes}, {Kasper}, {Kellogg}, {Klimchuk}, {Korreck},
  {Krasnoselskikh}, {Krucker}, {Laker}, {Larson}, {MacDowall}, {Maksimovic},
  {Malaspina}, {Martinez-Oliveros}, {McComas}, {Meyer-Vernet}, {Moncuquet},
  {Mozer}, {Phan}, {Pulupa}, {Raouafi}, {Salem}, {Stansby}, {Stevens}, {Szabo},
  {Velli}, {Woolley}, \& {Wygant}}]{2019Natur.576..237B}
{Bale}, S.~D., {Badman}, S.~T., {Bonnell}, J.~W., {et~al.} 2019, \nat, 576, 237

\bibitem[{{Boe} {et~al.}(2020){Boe}, {Habbal}, \&
  {Druckm{\"u}ller}}]{2020ApJ...895..123B}
{Boe}, B., {Habbal}, S., \& {Druckm{\"u}ller}, M. 2020, \apj, 895, 123

\bibitem[{{Brueckner} {et~al.}(1995){Brueckner}, {Howard}, {Koomen},
  {Korendyke}, {Michels}, {Moses}, {Socker}, {Dere}, {Lamy}, {Llebaria},
  {Bout}, {Schwenn}, {Simnett}, {Bedford}, \& {Eyles}}]{1995SoPh..162..357B}
{Brueckner}, G.~E., {Howard}, R.~A., {Koomen}, M.~J., {et~al.} 1995, \solphys,
  162, 357

\bibitem[{{Chen} {et~al.}(2021){Chen}, {Chandran}, {Woodham}, {Jones}, {Perez},
  {Bourouaine}, {Bowen}, {Klein}, {Moncuquet}, {Kasper}, \&
  {Bale}}]{2021A&A...650L...3C}
{Chen}, C.~H.~K., {Chandran}, B.~D.~G., {Woodham}, L.~D., {et~al.} 2021, \aap,
  650, L3

\bibitem[{{Chen} {et~al.}(2009){Chen}, {Li}, {Song}, {Shi}, {Feng}, \&
  {Xia}}]{2009ApJ...691.1936C}
{Chen}, Y., {Li}, X., {Song}, H.~Q., {et~al.} 2009, \apj, 691, 1936

\bibitem[{{Chitta} {et~al.}(2023){Chitta}, {Seaton}, {Downs}, {DeForest}, \&
  {Higginson}}]{2023NatAs...7..133C}
{Chitta}, L.~P., {Seaton}, D.~B., {Downs}, C., {DeForest}, C.~E., \&
  {Higginson}, A.~K. 2023, Nature Astronomy, 7, 133

\bibitem[{{Crooker} {et~al.}(2004){Crooker}, {Huang}, {Lamassa}, {Larson},
  {Kahler}, \& {Spence}}]{2004JGRA..109.3107C}
{Crooker}, N.~U., {Huang}, C.~L., {Lamassa}, S.~M., {et~al.} 2004, Journal of
  Geophysical Research (Space Physics), 109, A03107

\bibitem[{{Crooker} {et~al.}(1993){Crooker}, {Siscoe}, {Shodhan}, {Webb},
  {Gosling}, \& {Smith}}]{1993JGR....98.9371C}
{Crooker}, N.~U., {Siscoe}, G.~L., {Shodhan}, S., {et~al.} 1993, \jgr, 98, 9371

\bibitem[{{Domingo} {et~al.}(1995){Domingo}, {Fleck}, \&
  {Poland}}]{1995SoPh..162....1D}
{Domingo}, V., {Fleck}, B., \& {Poland}, A.~I. 1995, \solphys, 162, 1

\bibitem[{{Drake} {et~al.}(2021){Drake}, {Agapitov}, {Swisdak}, {Badman},
  {Bale}, {Horbury}, {Kasper}, {MacDowall}, {Mozer}, {Phan}, {Pulupa}, {Szabo},
  \& {Velli}}]{2021A&A...650A...2D}
{Drake}, J.~F., {Agapitov}, O., {Swisdak}, M., {et~al.} 2021, \aap, 650, A2

\bibitem[{{Einaudi} {et~al.}(1999){Einaudi}, {Boncinelli}, {Dahlburg}, \&
  {Karpen}}]{1999JGR...104..521E}
{Einaudi}, G., {Boncinelli}, P., {Dahlburg}, R.~B., \& {Karpen}, J.~T. 1999,
  \jgr, 104, 521

\bibitem[{{Eselevich} \& {Eselevich}(1999)}]{1999SoPh..188..299E}
{Eselevich}, V.~G. \& {Eselevich}, M.~V. 1999, \solphys, 188, 299

\bibitem[{{Fargette} {et~al.}(2021){Fargette}, {Lavraud}, {Rouillard},
  {R{\'e}ville}, {Dudok De Wit}, {Froment}, {Halekas}, {Phan}, {Malaspina},
  {Bale}, {Kasper}, {Louarn}, {Case}, {Korreck}, {Larson}, {Pulupa}, {Stevens},
  {Whittlesey}, \& {Berthomier}}]{2021ApJ...919...96F}
{Fargette}, N., {Lavraud}, B., {Rouillard}, A.~P., {et~al.} 2021, \apj, 919, 96

\bibitem[{{Grappin} {et~al.}(2000){Grappin}, {L{\'e}orat}, \&
  {Buttighoffer}}]{2000A&A...362..342G}
{Grappin}, R., {L{\'e}orat}, J., \& {Buttighoffer}, A. 2000, \aap, 362, 342

\bibitem[{{He} {et~al.}(2021){He}, {Zhu}, {Yang}, {Hou}, {Duan}, {Zhang}, \&
  {Wang}}]{2021ApJ...913L..14H}
{He}, J., {Zhu}, X., {Yang}, L., {et~al.} 2021, \apjl, 913, L14

\bibitem[{{Higginson} \& {Lynch}(2018)}]{2018ApJ...859....6H}
{Higginson}, A.~K. \& {Lynch}, B.~J. 2018, \apj, 859, 6

\bibitem[{{Hou} {et~al.}(2022){Hou}, {He}, {Duan}, {Li}, \&
  {Chen}}]{2022EGUGA..24.9673H}
{Hou}, C., {He}, J., {Duan}, D., {Li}, H., \& {Chen}, Y. 2022, in EGU General
  Assembly Conference Abstracts, EGU General Assembly Conference Abstracts,
  EGU22--9673

\bibitem[{{Hou} {et~al.}(2023){Hou}, {Zhu}, {Zhuo}, {He}, {Verscharen}, \&
  {Duan}}]{2023ApJ...950..157H}
{Hou}, C., {Zhu}, X., {Zhuo}, R., {et~al.} 2023, \apj, 950, 157

\bibitem[{{Karpen} {et~al.}(1998){Karpen}, {Antiochos}, {Richard DeVore}, \&
  {Golub}}]{1998ApJ...495..491K}
{Karpen}, J.~T., {Antiochos}, S.~K., {Richard DeVore}, C., \& {Golub}, L. 1998,
  \apj, 495, 491

\bibitem[{{Koutchmy} \& {Livshits}(1992)}]{1992SSRv...61..393K}
{Koutchmy}, S. \& {Livshits}, M. 1992, \ssr, 61, 393

\bibitem[{{Lee} {et~al.}(2021){Lee}, {Cho}, {An}, {Lee}, {Seough}, {Kim}, \&
  {Kumar}}]{2021ApJ...920L...6L}
{Lee}, J.-O., {Cho}, K.-S., {An}, J., {et~al.} 2021, \apjl, 920, L6

\bibitem[{{Lemen} {et~al.}(2012){Lemen}, {Title}, {Akin}, {Boerner}, {Chou},
  {Drake}, {Duncan}, {Edwards}, {Friedlaender}, {Heyman}, {Hurlburt}, {Katz},
  {Kushner}, {Levay}, {Lindgren}, {Mathur}, {McFeaters}, {Mitchell}, {Rehse},
  {Schrijver}, {Springer}, {Stern}, {Tarbell}, {Wuelser}, {Wolfson}, {Yanari},
  {Bookbinder}, {Cheimets}, {Caldwell}, {Deluca}, {Gates}, {Golub}, {Park},
  {Podgorski}, {Bush}, {Scherrer}, {Gummin}, {Smith}, {Auker}, {Jerram},
  {Pool}, {Soufli}, {Windt}, {Beardsley}, {Clapp}, {Lang}, \&
  {Waltham}}]{2012SoPh..275...17L}
{Lemen}, J.~R., {Title}, A.~M., {Akin}, D.~J., {et~al.} 2012, \solphys, 275, 17

\bibitem[{{Liang} {et~al.}(2021){Liang}, {Zank}, {Nakanotani}, \&
  {Zhao}}]{2021ApJ...917..110L}
{Liang}, H., {Zank}, G.~P., {Nakanotani}, M., \& {Zhao}, L.~L. 2021, \apj, 917,
  110

\bibitem[{{Liang} {et~al.}(2023){Liang}, {Qu}, {Hao}, {Xu}, \&
  {Zhong}}]{2023MNRAS.518.1776L}
{Liang}, Y., {Qu}, Z., {Hao}, L., {Xu}, Z., \& {Zhong}, Y. 2023, \mnras, 518,
  1776

\bibitem[{{Linker} \& {Mikic}(1995)}]{1995ApJ...438L..45L}
{Linker}, J.~A. \& {Mikic}, Z. 1995, \apjl, 438, L45

\bibitem[{{Lotova} {et~al.}(2000){Lotova}, {Obridko}, \&
  {Vladimirskii}}]{2000A&A...357.1051L}
{Lotova}, N.~A., {Obridko}, V.~N., \& {Vladimirskii}, K.~V. 2000, \aap, 357,
  1051

\bibitem[{{Lyu} {et~al.}(2023){Lyu}, {Wang}, {Li}, \&
  {Zhang}}]{2023A&A...672A.100L}
{Lyu}, S., {Wang}, Y., {Li}, X., \& {Zhang}, Q. 2023, \aap, 672, A100

\bibitem[{{Morgan} {et~al.}(2006){Morgan}, {Habbal}, \&
  {Woo}}]{2006SoPh..236..263M}
{Morgan}, H., {Habbal}, S.~R., \& {Woo}, R. 2006, \solphys, 236, 263

\bibitem[{{Mozer} {et~al.}(2021){Mozer}, {Bale}, {Bonnell}, {Drake}, {Hanson},
  \& {Mozer}}]{2021ApJ...919...60M}
{Mozer}, F.~S., {Bale}, S.~D., {Bonnell}, J.~W., {et~al.} 2021, \apj, 919, 60

\bibitem[{{M{\"u}ller} {et~al.}(2020){M{\"u}ller}, {St. Cyr}, {Zouganelis},
  {Gilbert}, {Marsden}, {Nieves-Chinchilla}, {Antonucci}, {Auch{\`e}re},
  {Berghmans}, {Horbury}, {Howard}, {Krucker}, {Maksimovic}, {Owen}, {Rochus},
  {Rodriguez-Pacheco}, {Romoli}, {Solanki}, {Bruno}, {Carlsson}, {Fludra},
  {Harra}, {Hassler}, {Livi}, {Louarn}, {Peter}, {Sch{\"u}hle}, {Teriaca}, {del
  Toro Iniesta}, {Wimmer-Schweingruber}, {Marsch}, {Velli}, {De Groof},
  {Walsh}, \& {Williams}}]{2020A&A...642A...1M}
{M{\"u}ller}, D., {St. Cyr}, O.~C., {Zouganelis}, I., {et~al.} 2020, \aap, 642,
  A1

\bibitem[{{Neugebauer} \& {Sterling}(2021)}]{2021ApJ...920L..31N}
{Neugebauer}, M. \& {Sterling}, A.~C. 2021, \apjl, 920, L31

\bibitem[{{Pasachoff} {et~al.}(2011){Pasachoff}, {Ru{\v{s}}in},
  {Druckm{\"u}llerov{\'a}}, {Saniga}, {Lu}, {Malamut}, {Seaton}, {Golub},
  {Engell}, {Hill}, \& {Lucas}}]{2011ApJ...734..114P}
{Pasachoff}, J.~M., {Ru{\v{s}}in}, V., {Druckm{\"u}llerov{\'a}}, H., {et~al.}
  2011, \apj, 734, 114

\bibitem[{{Pasachoff} {et~al.}(2015){Pasachoff}, {Ru{\v{s}}in}, {Saniga},
  {Babcock}, {Lu}, {Davis}, {Dantowitz}, {Gaintatzis}, {Seiradakis},
  {Voulgaris}, {Seaton}, \& {Shiota}}]{2015ApJ...800...90P}
{Pasachoff}, J.~M., {Ru{\v{s}}in}, V., {Saniga}, M., {et~al.} 2015, \apj, 800,
  90

\bibitem[{{Pellegrin-Frachon} {et~al.}(2023){Pellegrin-Frachon}, {Masson},
  {Pariat}, {Wyper}, \& {DeVore}}]{2023A&A...675A..55P}
{Pellegrin-Frachon}, T., {Masson}, S., {Pariat}, {\'E}., {Wyper}, P.~F., \&
  {DeVore}, C.~R. 2023, \aap, 675, A55

\bibitem[{{Pesnell} {et~al.}(2012){Pesnell}, {Thompson}, \&
  {Chamberlin}}]{2012SoPh..275....3P}
{Pesnell}, W.~D., {Thompson}, B.~J., \& {Chamberlin}, P.~C. 2012, \solphys,
  275, 3

\bibitem[{{Phan} {et~al.}(2021){Phan}, {Lavraud}, {Halekas}, {{\O}ieroset},
  {Drake}, {Eastwood}, {Shay}, {Pyakurel}, {Bale}, {Larson}, {Livi},
  {Whittlesey}, {Rahmati}, {Pulupa}, {McManus}, {Verniero}, {Bonnell},
  {Schwadron}, {Stevens}, {Case}, {Kasper}, {MacDowall}, {Szabo}, {Koval},
  {Korreck}, {Dudok de Wit}, {Malaspina}, {Goetz}, \&
  {Harvey}}]{2021A&A...650A..13P}
{Phan}, T.~D., {Lavraud}, B., {Halekas}, J.~S., {et~al.} 2021, \aap, 650, A13

\bibitem[{{Poirier} {et~al.}(2020){Poirier}, {Kouloumvakos}, {Rouillard},
  {Pinto}, {Vourlidas}, {Stenborg}, {Valette}, {Howard}, {Hess}, {Thernisien},
  {Rich}, {Griton}, {Indurain}, {Raouafi}, {Lavarra}, \&
  {R{\'e}ville}}]{2020ApJS..246...60P}
{Poirier}, N., {Kouloumvakos}, A., {Rouillard}, A.~P., {et~al.} 2020, \apjs,
  246, 60

\bibitem[{{Raouafi} {et~al.}(2023){Raouafi}, {Stenborg}, {Seaton}, {Wang},
  {Wang}, {DeForest}, {Bale}, {Drake}, {Uritsky}, {Karpen}, {DeVore},
  {Sterling}, {Horbury}, {Harra}, {Bourouaine}, {Kasper}, {Kumar}, {Phan}, \&
  {Velli}}]{2023ApJ...945...28R}
{Raouafi}, N.~E., {Stenborg}, G., {Seaton}, D.~B., {et~al.} 2023, \apj, 945, 28

\bibitem[{{Raymond} {et~al.}(1997){Raymond}, {Kohl}, {Noci}, {Antonucci},
  {Tondello}, {Huber}, {Gardner}, {Nicolosi}, {Fineschi}, {Romoli}, {Spadaro},
  {Siegmund}, {Benna}, {Ciaravella}, {Cranmer}, {Giordano}, {Karovska},
  {Martin}, {Michels}, {Modigliani}, {Naletto}, {Panasyuk}, {Pernechele},
  {Poletto}, {Smith}, {Suleiman}, \& {Strachan}}]{1997SoPh..175..645R}
{Raymond}, J.~C., {Kohl}, J.~L., {Noci}, G., {et~al.} 1997, \solphys, 175, 645

\bibitem[{{Rochus} {et~al.}(2020){Rochus}, {Auch{\`e}re}, {Berghmans}, {Harra},
  {Schmutz}, {Sch{\"u}hle}, {Addison}, {Appourchaux}, {Aznar Cuadrado},
  {Baker}, {Barbay}, {Bates}, {BenMoussa}, {Bergmann}, {Beurthe}, {Borgo},
  {Bonte}, {Bouzit}, {Bradley}, {B{\"u}chel}, {Buchlin}, {B{\"u}chner},
  {Cab{\'e}}, {Cadiergues}, {Chaigneau}, {Chares}, {Choque Cortez}, {Coker},
  {Condamin}, {Coumar}, {Curdt}, {Cutler}, {Davies}, {Davison}, {Defise}, {Del
  Zanna}, {Delmotte}, {Delouille}, {Dolla}, {Dumesnil}, {D{\"u}rig}, {Enge},
  {Fran{\c{c}}ois}, {Fourmond}, {Gillis}, {Giordanengo}, {Gissot}, {Green},
  {Guerreiro}, {Guilbaud}, {Gyo}, {Haberreiter}, {Hafiz}, {Hailey}, {Halain},
  {Hansotte}, {Hecquet}, {Heerlein}, {Hellin}, {Hemsley}, {Hermans}, {Hervier},
  {Hochedez}, {Houbrechts}, {Ihsan}, {Jacques}, {J{\'e}r{\^o}me}, {Jones},
  {Kahle}, {Kennedy}, {Klaproth}, {Kolleck}, {Koller}, {Kotsialos},
  {Kraaikamp}, {Langer}, {Lawrenson}, {Le Clech'}, {Lenaerts}, {Liebecq},
  {Linder}, {Long}, {Mampaey}, {Markiewicz-Innes}, {Marquet}, {Marsch},
  {Matthews}, {Mazy}, {Mazzoli}, {Meining}, {Meltchakov}, {Mercier}, {Meyer},
  {Monecke}, {Monfort}, {Morinaud}, {Moron}, {Mountney}, {M{\"u}ller},
  {Nicula}, {Parenti}, {Peter}, {Pfiffner}, {Philippon}, {Phillips},
  {Plesseria}, {Pylyser}, {Rabecki}, {Ravet-Krill}, {Rebellato}, {Renotte},
  {Rodriguez}, {Roose}, {Rosin}, {Rossi}, {Roth}, {Rouesnel}, {Roulliay},
  {Rousseau}, {Ruane}, {Scanlan}, {Schlatter}, {Seaton}, {Silliman}, {Smit},
  {Smith}, {Solanki}, {Spescha}, {Spencer}, {Stegen}, {Stockman}, {Szwec},
  {Tamiatto}, {Tandy}, {Teriaca}, {Theobald}, {Tychon}, {van Driel-Gesztelyi},
  {Verbeeck}, {Vial}, {Werner}, {West}, {Westwood}, {Wiegelmann}, {Willis},
  {Winter}, {Zerr}, {Zhang}, \& {Zhukov}}]{2020A&A...642A...8R}
{Rochus}, P., {Auch{\`e}re}, F., {Berghmans}, D., {et~al.} 2020, \aap, 642, A8

\bibitem[{{Romoli} {et~al.}(2021){Romoli}, {Antonucci}, {Andretta}, {Capuano},
  {Da Deppo}, {De Leo}, {Downs}, {Fineschi}, {Heinzel}, {Landini},
  {Liberatore}, {Naletto}, {Nicolini}, {Pancrazzi}, {Sasso}, {Spadaro},
  {Susino}, {Telloni}, {Teriaca}, {Uslenghi}, {Wang}, {Bemporad}, {Capobianco},
  {Casti}, {Fabi}, {Frassati}, {Frassetto}, {Giordano}, {Grimani}, {Jerse},
  {Magli}, {Massone}, {Messerotti}, {Moses}, {Pelizzo}, {Romano},
  {Sch{\"u}hle}, {Slemer}, {Stangalini}, {Straus}, {Volpicelli}, {Zangrilli},
  {Zuppella}, {Abbo}, {Auch{\`e}re}, {Aznar Cuadrado}, {Berlicki}, {Bruno},
  {Ciaravella}, {D'Amicis}, {Lamy}, {Lanzafame}, {Malvezzi}, {Nicolosi},
  {Nistic{\`o}}, {Peter}, {Plainaki}, {Poletto}, {Reale}, {Solanki},
  {Strachan}, {Tondello}, {Tsinganos}, {Velli}, {Ventura}, {Vial}, {Woch}, \&
  {Zimbardo}}]{2021A&A...656A..32R}
{Romoli}, M., {Antonucci}, E., {Andretta}, V., {et~al.} 2021, \aap, 656, A32

\bibitem[{{Rouillard} {et~al.}(2020){Rouillard}, {Kouloumvakos}, {Vourlidas},
  {Kasper}, {Bale}, {Raouafi}, {Lavraud}, {Howard}, {Stenborg}, {Stevens},
  {Poirier}, {Davies}, {Hess}, {Higginson}, {Lavarra}, {Viall}, {Korreck},
  {Pinto}, {Griton}, {R{\'e}ville}, {Louarn}, {Wu}, {Dalmasse}, {G{\'e}not},
  {Case}, {Whittlesey}, {Larson}, {Halekas}, {Livi}, {Goetz}, {Harvey},
  {MacDowall}, {Malaspina}, {Pulupa}, {Bonnell}, {de Witt}, \&
  {Penou}}]{2020ApJS..246...37R}
{Rouillard}, A.~P., {Kouloumvakos}, A., {Vourlidas}, A., {et~al.} 2020, \apjs,
  246, 37

\bibitem[{{Saito} \& {Tandberg-Hanssen}(1973)}]{1973SoPh...31..105S}
{Saito}, K. \& {Tandberg-Hanssen}, E. 1973, \solphys, 31, 105

\bibitem[{{Seaton} \& {Darnel}(2018)}]{2018ApJ...852L...9S}
{Seaton}, D.~B. \& {Darnel}, J.~M. 2018, \apjl, 852, L9

\bibitem[{{Seaton} {et~al.}(2021){Seaton}, {Hughes}, {Tadikonda}, {Caspi},
  {DeForest}, {Krimchansky}, {Hurlburt}, {Seguin}, \&
  {Slater}}]{2021NatAs...5.1029S}
{Seaton}, D.~B., {Hughes}, J.~M., {Tadikonda}, S.~K., {et~al.} 2021, Nature
  Astronomy, 5, 1029

\bibitem[{{Sheeley} {et~al.}(2009){Sheeley}, {Lee}, {Casto}, {Wang}, \&
  {Rich}}]{2009ApJ...694.1471S}
{Sheeley}, N.~R., J., {Lee}, D.~D.~H., {Casto}, K.~P., {Wang}, Y.~M., \&
  {Rich}, N.~B. 2009, \apj, 694, 1471

\bibitem[{{Sheeley} \& {Wang}(2007)}]{2007ApJ...655.1142S}
{Sheeley}, N.~R., J. \& {Wang}, Y.~M. 2007, \apj, 655, 1142

\bibitem[{{Sheeley} {et~al.}(1999){Sheeley}, {Walters}, {Wang}, \&
  {Howard}}]{1999JGR...10424739S}
{Sheeley}, N.~R., {Walters}, J.~H., {Wang}, Y.~M., \& {Howard}, R.~A. 1999,
  \jgr, 104, 24739

\bibitem[{{Sheeley} {et~al.}(1997){Sheeley}, {Wang}, {Hawley}, {Brueckner},
  {Dere}, {Howard}, {Koomen}, {Korendyke}, {Michels}, {Paswaters}, {Socker},
  {St. Cyr}, {Wang}, {Lamy}, {Llebaria}, {Schwenn}, {Simnett}, {Plunkett}, \&
  {Biesecker}}]{1997ApJ...484..472S}
{Sheeley}, N.~R., {Wang}, Y.~M., {Hawley}, S.~H., {et~al.} 1997, \apj, 484, 472

\bibitem[{{Song} {et~al.}(2009){Song}, {Chen}, {Liu}, {Feng}, \&
  {Xia}}]{2009SoPh..258..129S}
{Song}, H.~Q., {Chen}, Y., {Liu}, K., {Feng}, S.~W., \& {Xia}, L.~D. 2009,
  \solphys, 258, 129

\bibitem[{{Song} {et~al.}(2012){Song}, {Kong}, {Chen}, {Li}, {Li}, {Feng}, \&
  {Xia}}]{2012SoPh..276..261S}
{Song}, H.~Q., {Kong}, X.~L., {Chen}, Y., {et~al.} 2012, \solphys, 276, 261

\bibitem[{{Spadaro} {et~al.}(2007){Spadaro}, {Susino}, {Ventura}, {Vourlidas},
  \& {Landi}}]{2007A&A...475..707S}
{Spadaro}, D., {Susino}, R., {Ventura}, R., {Vourlidas}, A., \& {Landi}, E.
  2007, \aap, 475, 707

\bibitem[{{Squire} {et~al.}(2020){Squire}, {Chandran}, \&
  {Meyrand}}]{2020ApJ...891L...2S}
{Squire}, J., {Chandran}, B.~D.~G., \& {Meyrand}, R. 2020, \apjl, 891, L2

\bibitem[{{Strachan} {et~al.}(2002){Strachan}, {Suleiman}, {Panasyuk},
  {Biesecker}, \& {Kohl}}]{2002ApJ...571.1008S}
{Strachan}, L., {Suleiman}, R., {Panasyuk}, A.~V., {Biesecker}, D.~A., \&
  {Kohl}, J.~L. 2002, \apj, 571, 1008

\bibitem[{{Sturrock} \& {Smith}(1968)}]{1968SoPh....5...87S}
{Sturrock}, P.~A. \& {Smith}, S.~M. 1968, \solphys, 5, 87

\bibitem[{{Suess} {et~al.}(2009){Suess}, {Ko}, {von Steiger}, \&
  {Moore}}]{2009JGRA..114.4103S}
{Suess}, S.~T., {Ko}, Y.~K., {von Steiger}, R., \& {Moore}, R.~L. 2009, Journal
  of Geophysical Research (Space Physics), 114, A04103

\bibitem[{{Telloni} {et~al.}(2022){Telloni}, {Zank}, {Stangalini}, {Downs},
  {Liang}, {Nakanotani}, {Andretta}, {Antonucci}, {Sorriso-Valvo}, {Adhikari},
  {Zhao}, {Marino}, {Susino}, {Grimani}, {Fabi}, {D'Amicis}, {Perrone},
  {Bruno}, {Carbone}, {Mancuso}, {Romoli}, {Deppo}, {Fineschi}, {Heinzel},
  {Moses}, {Naletto}, {Nicolini}, {Spadaro}, {Teriaca}, {Frassati}, {Jerse},
  {Landini}, {Pancrazzi}, {Russano}, {Sasso}, {Biondo}, {Burtovoi}, {Capuano},
  {Casini}, {Casti}, {Chioetto}, {De Leo}, {Giarrusso}, {Liberatore},
  {Berghmans}, {Auch{\`e}re}, {Cuadrado}, {Chitta}, {Harra}, {Kraaikamp},
  {Long}, {Mandal}, {Parenti}, {Pelouze}, {Peter}, {Rodriguez}, {Sch{\"u}hle},
  {Schwanitz}, {Smith}, {Verbeeck}, \& {Zhukov}}]{2022ApJ...936L..25T}
{Telloni}, D., {Zank}, G.~P., {Stangalini}, M., {et~al.} 2022, \apjl, 936, L25

\bibitem[{{Tenerani} {et~al.}(2020){Tenerani}, {Velli}, {Matteini},
  {R{\'e}ville}, {Shi}, {Bale}, {Kasper}, {Bonnell}, {Case}, {de Wit}, {Goetz},
  {Harvey}, {Klein}, {Korreck}, {Larson}, {Livi}, {MacDowall}, {Malaspina},
  {Pulupa}, {Stevens}, \& {Whittlesey}}]{2020ApJS..246...32T}
{Tenerani}, A., {Velli}, M., {Matteini}, L., {et~al.} 2020, \apjs, 246, 32

\bibitem[{{Thernisien} \& {Howard}(2006)}]{2006ApJ...642..523T}
{Thernisien}, A.~F. \& {Howard}, R.~A. 2006, \apj, 642, 523

\bibitem[{{Upendran} \& {Tripathi}(2022)}]{2022ApJ...926..138U}
{Upendran}, V. \& {Tripathi}, D. 2022, \apj, 926, 138

\bibitem[{{Uzzo} {et~al.}(2003){Uzzo}, {Ko}, {Raymond}, {Wurz}, \&
  {Ipavich}}]{2003ApJ...585.1062U}
{Uzzo}, M., {Ko}, Y.~K., {Raymond}, J.~C., {Wurz}, P., \& {Ipavich}, F.~M.
  2003, \apj, 585, 1062

\bibitem[{{Vasudevan} {et~al.}(2019){Vasudevan}, {Shing}, {Mathur}, {Edwards},
  {Shaw}, {Seaton}, \& {Darnel}}]{2019SPIE11180E..7PV}
{Vasudevan}, G., {Shing}, L., {Mathur}, D., {et~al.} 2019, in Society of
  Photo-Optical Instrumentation Engineers (SPIE) Conference Series, Vol. 11180,
  International Conference on Space Optics; ICSO 2018, ed. Z.~{Sodnik},
  N.~{Karafolas}, \& B.~{Cugny}, 111807P

\bibitem[{{Ventura} {et~al.}(2023){Ventura}, {Antonucci}, {Downs}, {Romano},
  {Susino}, {Spadaro}, {Telloni}, {Guglielmino}, {Capuano}, {Andretta},
  {Landini}, {Jerse}, {Nicolini}, {Pancrazzi}, {Sasso}, {Da Deppo}, {Fineschi},
  {Grimani}, {Heinzel}, {Moses}, {Naletto}, {Romoli}, {Stangalini}, {Teriaca},
  \& {Uslenghi}}]{2023A&A...675A.170V}
{Ventura}, R., {Antonucci}, E., {Downs}, C., {et~al.} 2023, \aap, 675, A170

\bibitem[{{Ventura} {et~al.}(2005){Ventura}, {Spadaro}, {Cimino}, \&
  {Romoli}}]{2005A&A...430..701V}
{Ventura}, R., {Spadaro}, D., {Cimino}, G., \& {Romoli}, M. 2005, \aap, 430,
  701

\bibitem[{{Wang}(2012)}]{2012SSRv..172..123W}
{Wang}, Y.~M. 2012, \ssr, 172, 123

\bibitem[{{Wang} \& {Hess}(2018)}]{2018ApJ...859..135W}
{Wang}, Y.~M. \& {Hess}, P. 2018, \apj, 859, 135

\bibitem[{{Wang} \& {Sheeley}(1992)}]{1992ApJ...392..310W}
{Wang}, Y.~M. \& {Sheeley}, N.~R., J. 1992, \apj, 392, 310

\bibitem[{{Wang} \& {Sheeley}(2006)}]{2006ApJ...650.1172W}
{Wang}, Y.~M. \& {Sheeley}, N.~R., J. 2006, \apj, 650, 1172

\bibitem[{{Wang} {et~al.}(2000){Wang}, {Sheeley}, \&
  {Rich}}]{2000GeoRL..27..149W}
{Wang}, Y.~M., {Sheeley}, N.~R., J., \& {Rich}, N.~B. 2000, \grl, 27, 149

\bibitem[{{Wang} {et~al.}(2007){Wang}, {Sheeley}, \&
  {Rich}}]{2007ApJ...658.1340W}
{Wang}, Y.~M., {Sheeley}, N.~R., J., \& {Rich}, N.~B. 2007, \apj, 658, 1340

\bibitem[{{Wang} {et~al.}(1998){Wang}, {Sheeley}, {Walters}, {Brueckner},
  {Howard}, {Michels}, {Lamy}, {Schwenn}, \& {Simnett}}]{1998ApJ...498L.165W}
{Wang}, Y.~M., {Sheeley}, N.~R., J., {Walters}, J.~H., {et~al.} 1998, \apjl,
  498, L165

\bibitem[{{West} {et~al.}(2023){West}, {Seaton}, {Wexler}, {Raymond}, {Del
  Zanna}, {Rivera}, {Kobelski}, {Chen}, {DeForest}, {Golub}, {Caspi}, {Gilly},
  {Kooi}, {Meyer}, {Alterman}, {Alzate}, {Andretta}, {Auch{\`e}re}, {Banerjee},
  {Berghmans}, {Chamberlin}, {Chitta}, {Downs}, {Giordano}, {Harra},
  {Higginson}, {Howard}, {Kumar}, {Mason}, {Mason}, {Morton}, {Nykyri},
  {Patel}, {Rachmeler}, {Reardon}, {Reeves}, {Savage}, {Thompson}, {Van
  Kooten}, {Viall}, {Vourlidas}, \& {Zhukov}}]{2023SoPh..298...78W}
{West}, M.~J., {Seaton}, D.~B., {Wexler}, D.~B., {et~al.} 2023, \solphys, 298,
  78

\bibitem[{{Winterhalter} {et~al.}(1994){Winterhalter}, {Smith}, {Burton},
  {Murphy}, \& {McComas}}]{1994JGR....99.6667W}
{Winterhalter}, D., {Smith}, E.~J., {Burton}, M.~E., {Murphy}, N., \&
  {McComas}, D.~J. 1994, \jgr, 99, 6667

\bibitem[{{Wyper} {et~al.}(2022){Wyper}, {DeVore}, {Antiochos}, {Pontin},
  {Higginson}, {Scott}, {Masson}, \& {Pelegrin-Frachon}}]{2022ApJ...941L..29W}
{Wyper}, P.~F., {DeVore}, C.~R., {Antiochos}, S.~K., {et~al.} 2022, \apjl, 941,
  L29

\bibitem[{{Zank} {et~al.}(2020){Zank}, {Nakanotani}, {Zhao}, {Adhikari}, \&
  {Kasper}}]{2020ApJ...903....1Z}
{Zank}, G.~P., {Nakanotani}, M., {Zhao}, L.~L., {Adhikari}, L., \& {Kasper}, J.
  2020, \apj, 903, 1

\bibitem[{{Zhao} {et~al.}(2021){Zhao}, {Zank}, {Hu}, {Telloni}, {Chen},
  {Adhikari}, {Nakanotani}, {Kasper}, {Huang}, {Bale}, {Korreck}, {Case},
  {Stevens}, {Bonnell}, {Dudok de Wit}, {Goetz}, {Harvey}, {MacDowall},
  {Malaspina}, {Pulupa}, {Larson}, {Livi}, {Whittlesey}, {Klein}, \&
  {Raouafi}}]{2021A&A...650A..12Z}
{Zhao}, L.~L., {Zank}, G.~P., {Hu}, Q., {et~al.} 2021, \aap, 650, A12

\end{thebibliography}
\bibliographystyle{aa}


%
%
\end{document}